\documentclass[letterpaper]{article} 
\usepackage{aaai2026}  
\usepackage{times}  
\usepackage{helvet}  
\usepackage{courier}  
\usepackage[hyphens]{url}  
\usepackage{graphicx} 
\usepackage{natbib}  
\usepackage{caption} 
\usepackage{algorithm}
\usepackage{algorithmic}
\usepackage{amsmath}
\usepackage{mathtools}
\DeclareMathOperator*{\argmin}{arg\,min}
\usepackage{newfloat}
\usepackage{listings}
\DeclareCaptionStyle{ruled}{labelfont=normalfont,labelsep=colon,strut=off} 
\floatstyle{ruled}
\newfloat{listing}{tb}{lst}{}
\floatname{listing}{Listing}
\title{Deep RL for Fast Long-Horizon Operations Scheduling on NASA’s Carruthers Geocorona Observatory Mission}
\author{
    Alex M. Zhang\textsuperscript{\rm 1},
    Jackson Craig\textsuperscript{\rm 1},
    Lara Waldrop\textsuperscript{\rm 1}
}
\affiliations{
    \textsuperscript{\rm 1}Department of Electrical and Computer Engineering, University of Illinois Urbana-Champaign, IL, USA \\
    \{alexmz2, jc168, lwaldrop\}@illinois.edu
}

\begin{document}

\maketitle

\begin{abstract}
Spacecraft operations scheduling is a highly constrained, long-horizon combinatorial optimization problem that traditionally relies on heuristics, constraint programming, or manual planning. We present a scalable deep reinforcement learning framework developed and deployed for NASA's Carruthers Geocorona Observatory mission. Our framework introduces a macro-action abstraction known as activity blocks coupled with dynamic action-masking to navigate the intractably large search space and strictly enforce complex power, thermal, and instrument constraints. The resulting architecture generates globally feasible schedules with overwhelming probability, establishes operational trust, and executes a full training cycle in under six hours, circumventing the need for policy robustness by enabling rapid, on-demand retraining. Further, resulting schedules outperform baseline heuristics in scheduled science quality. The deep reinforcement learning framework was deployed as the default operational scheduler for the Carruthers Geocorona Observatory mission from the outset of the mission, demonstrating that deep reinforcement learning can be trusted for real spacecraft operations under complex, evolving constraints.
\end{abstract}

\section{Introduction}

Science-driven space missions routinely face long-horizon combinatorial scheduling problems, where each observation sequence must satisfy power, thermal, pointing, instrument, and data-volume constraints while maximizing scientific return. NASA's Carruthers Geocorona Observatory mission, launched in September 2025, is no exception: the scheduling problem faced by the Carruthers mission covers a long horizon and is particularly challenging both in scale and cost function complexity. Finally, like most space missions, the mission's scheduling system is required to be highly reactive or flexible in the case of anomalies or science Targets of Opportunity (ToO). These factors place the scheduling problem far beyond the reach of manual planning, rule-based heuristics \cite{rumford2003byhand1}, mixed-integer programming \cite{sabol2021dsnscheduling}, constraint programming \cite{barreiro2012europascheduler}, graph search \cite{jacquet2024earth}, Planning Domain Definition Language \cite{aeronautiques1998pddl}, or classical reinforcement learning \cite{tipaldi2017surveyclassicrl}.

Deep reinforcement learning (DRL) has recently shown promise for long-horizon scheduling by sequentially constructing schedules using neural-network–based policies \cite{herrmann2023comparative}. However, such approaches are faced with three major issues:
\begin{enumerate}
    \item Operational Trust: Does the system provide sufficient plan transparency for stakeholders to easily verify constraint adherence and validate scientific objectives?
    \item Agility: Is the trained policy robust to constraint and/or science priority changes?
    \item Strict Feasibility: Are the schedules generated by the approach always feasible?
\end{enumerate}

In this work, we present a scalable DRL framework and apply it to the scheduling problem faced by the Carruthers mission. Our framework combines activity blocks, a macro-action abstraction that compresses sequences of atomic decisions into high-level scheduling blocks, with action-masking, which enforces feasibility in large, heavily-constrained action spaces and stabilizes training. Together, these design choices reduce the effective search space by thousands of orders of magnitude, make scheduling tractable, and enable real-world deployment. The resulting schedules are consistently feasible, provide sufficient plan transparency to mission stakeholders, and are produced in under six hours of computation, thereby allowing for complete retraining under changing constraints. Crucially, this framework has served as the default operational scheduler for the Carruthers mission from launch, demonstrating that DRL can be trusted for real spacecraft operations under evolving constraints.

\section{Carruthers Geocorona Observatory}

The Carruthers Geocorona Observatory is NASA’s first space mission dedicated to investigating the fundamental nature of Earth’s exosphere. The mission’s primary payload, the GeoCoronal Imager (GCI), consists of two co-aligned broadband photometric imagers (channels), each equipped with an independent 6-position filter wheel \cite{rider2024alignment}. The mission operates in an orbit around the Earth–Sun L1 Lagrange point, in deep space.

Accurate spatial reconstruction of exospheric hydrogen density depends critically on absolute calibration, i.e. quantifying the instrument's exact response to incident photons. Nominal responsivities were derived during pre-launch testing, but on-orbit calibration is required to characterize potential responsivity degradations due to thermal variability or contaminant deposition. The nominal operations plan is to observe a subset of a catalog of visible, stable, and well-characterized calibration stars, derived from the extensive measurement archives of past missions such as the International Ultraviolet Explorer (IUE) \cite{klinglesmith1979iue_image_processing_system}, the Hubble Space Telescope (HST) \cite{bohlin2019hubble_flux_cal}, and others \cite{Zhang26sensitivity}.

However, stars are not monochromatic, which implies that the recovery of the instrument's continuous responsivity $r_{c, f}(\lambda)$ on channel $c$ and filter $f$ as a function of wavelength $\lambda$ requires solving an underdetermined linear system. Let $\vec{S}_{\text{stars}}$ be the vector of measured instrument responses across a set of stellar observations, $\boldsymbol{L}$ be the matrix of known stellar fluxes, and $\vec{w}$ represent some zero-mean Gaussian measurement noise with diagonal covariance. The forward model is given by $\vec{S}_{\text{stars}} = \boldsymbol{L}\vec{r}_{c, f}+\vec{w}$, where $\vec{r}_{c, f}$ is the discretized version of $r_{c, f}(\lambda)$. To recover responsivity, we minimize the error between observed and expected signals, regularized by a prior $\vec{p}_{c, f}$ measured during pre-launch testing:
\begin{equation}
\label{eq:abs_resp_min}
    \vec{r}_{c, f} = \argmin_{\vec{r}_{c, f}} |\vec{S}_{\text{stars}} - \boldsymbol{L} \vec{r}_{c, f}|_2^2 + \gamma |\vec{p}_{c, f} \circ \boldsymbol{D}(\vec{r}_{c, f} - \vec{p}_{c, f})|_2^2
\end{equation}
Here, $\boldsymbol{D}$ is the first difference matrix such that $(\boldsymbol{D}x)_i = x_i - x_{i-1}$, $\gamma$ is the regularization hyperparameter, and $\circ$ denotes the vector Hadamard product.

Due to operational time constraints, observing the complete catalog of candidate stars across all 10 channel/filter combinations with sufficient integration time for a desirable signal-to-noise ratio (SNR) is infeasible. Consequently, a subset of targets that maximizes responsivity-retrieval accuracy must be identified. However, the target selection problem cannot be decoupled from the mission's operational constraints; the theoretically optimal subset for calibration may yield an operationally infeasible schedule. Addressing this tight coupling between scientific accuracy maximization and strict schedule feasibility forms the central problem addressed in this paper.

\section{Scheduling Problem Formulation}
\label{sec:sched_prob}

In this section, we formally define the scheduling problem and the constraints.

\subsection{Definitions and Notation}

The scheduling horizon spans from $t_0$ to $t_f$, comprising $n_d$ days and $n_m$ four-week periods. Specific durations are denoted by $t_{\text{[duration]}}$ and evaluated in seconds (e.g., $t_{\text{1hr}} = 3600$, $t_{\text{4wk}} = 2419200$). Let $\mathcal{I}$ denote the set of all scheduled images and $\mathcal{B}$ the set of all blackout regions. These blackout windows are reserved for momentum dumps, communications, or spacecraft maneuvers.

An image $i \in \mathcal{I}$ is defined by the tuple $(s_i, d_i, f_i, c_i, \tau_i)$, representing its start time, duration, filter, channel, and target, respectively. Its end time is $e_i = s_i + d_i$. A blackout region $b \in \mathcal{B}$ is similarly defined by $(s_b, d_b, \tau_b)$, where the target $\tau_b$ is always Earth.

To formally express temporal intersections, we define the overlap duration operator $\Omega$ for any two time intervals $[s_x, e_x]$ and $[s_y, e_y]$:
\begin{equation*}
    \Omega(s_x, e_x, s_y, e_y) = \max(0, \min(e_x, e_y) - \max(s_x, s_y))
\end{equation*}

For notational convenience, let the function $s_c(x)$ return the start time of the next scheduled image or blackout on channel $c$ immediately following event $x$. If there is idle time between consecutive images, then $s_c(x) \neq e_x$. Let $\hat{s}_i(t) = \max(s_i, t)$ and $\hat{e}_i(t) = \min(s_c(i), t)$ denote the start and end times of event $i$ bounded by a time $t$.

\begin{figure}[t]
\centering
\includegraphics[width=0.9\columnwidth]{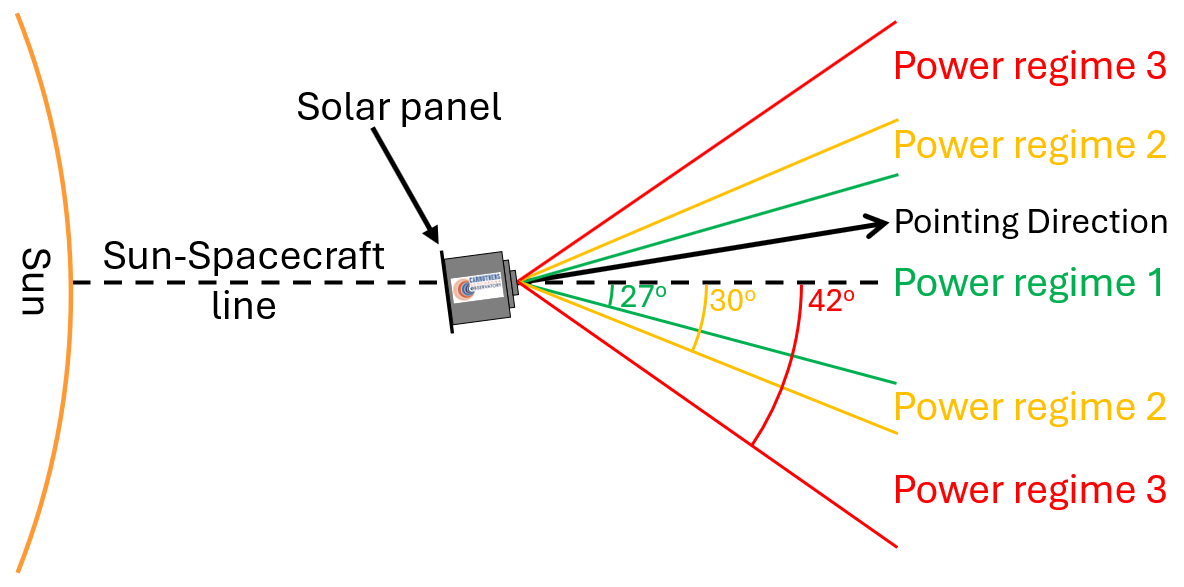}
\caption{Power regime definition.}
\label{fig:power_regime_def}
\end{figure}

Power generation and thermal limits are determined by the angle between the solar-panel normal and the Sun–spacecraft line when the spacecraft is pointed at a target $\tau$ (see Figure \ref{fig:power_regime_def}), with larger angles producing less power and causing more heating. Let $p_k(\tau, s, e)$ denote the continuous time, in seconds, spent in Power Regime $k$ given a target $\tau$ over the interval $[s, e]$. Instrument pointing is ``sticky''; the target $\tau$ remains constant after an image completes until the next event begins. This is a conservative approximation for power usage during complex spacecraft slews.

\subsection{Instrument Constraints}
The instrument constraints arise from the characteristics of the instrument itself.

\noindent \textbf{[INST-1]:} Images on the same channel cannot overlap in time.
\begin{equation*}
    \forall i\neq j: c_i = c_j \implies \Omega(s_i, e_i, s_j, e_j) = 0
\end{equation*}

\noindent \textbf{[INST-2]:} All images must execute completely within the global scheduling window.
\begin{equation*}
    \forall i: \Omega(s_i, e_i, t_0, t_f) = d_i
\end{equation*}

\noindent \textbf{[INST-3]:} Image durations must be less than 60 minutes.
\begin{equation*}
    \forall i: d_i \leq t_{\text{1hr}}
\end{equation*}

\noindent \textbf{[INST-4]:} The two instrument channels are physically co-aligned; concurrent images must observe the same target.
\begin{equation*}
    \forall i \neq j: c_i \neq c_j \land \Omega(s_i, e_i, s_j, e_j) > 0 \implies \tau_i = \tau_j
\end{equation*}

\noindent \textbf{[INST-5]:} Consecutive images utilizing different filters require a 30-second idle period for wheel rotation.
\begin{equation*}
    \forall i \neq j: c_i = c_j \land f_i \neq f_j \implies \max(s_j - e_i, s_i - e_j) \geq 30
\end{equation*}

\noindent \textbf{[INST-6]:} Consecutive images with different targets require a 10-minute idle time for spacecraft slew and settle.
\begin{equation*}
    \forall i \neq j: c_i = c_j \land \tau_i \neq \tau_j \implies \max(s_j - e_i, s_i - e_j) \geq 600
\end{equation*}

\noindent \textbf{[INST-7]:} Consecutive images must span a combined temporal block of at least 10 minutes, inclusive of idle time.
\begin{equation*}
    \forall i \neq j: c_i = c_j \implies s_c(i) - s_j \geq 600 \lor s_c(j) - s_i \geq 600
\end{equation*}

\noindent \textbf{[INST-8]:} No imaging is permitted during specified blackout windows.
\begin{equation*}
    \forall b \in \mathcal{B}, i \in \mathcal{I}: \Omega(s_b, e_b, s_i, e_i) = 0
\end{equation*}

\subsection{Power and Thermal Constraints}
These constraints are derived from pre-launch models.

\noindent \textbf{[POWER-1]:} An image exposure must remain entirely within the union of Power Regimes 1, 2, and 3.
\begin{equation*}
    \forall i: \sum_{k = 1}^3 p_k(\tau_i, s_i, e_i) = d_i
\end{equation*}

\noindent \textbf{[POWER-2]:} Continuous imaging within the power-negative Regimes 2 and 3 cannot exceed 8 hours without an intervening power-positive Regime 1 event.
\begin{equation*}
\begin{aligned}
    \forall i, j: & c_i = c_j \land s_i < e_j \land e_j - s_i > t_{\text{8hr}} \\
    & \implies \exists m\in(\mathcal{I}\cup \mathcal{B}) : c_m = c_i {}\\
    &\quad \land \Omega(s_i, e_j, s_m, e_m) > 0 \land p_1(\tau_m, s_m, e_m) > 0
\end{aligned}
\end{equation*}

\noindent \textbf{[POWER-3]:} Imaging within Power Regime 2 is restricted to a maximum of 8 hours per 24-hour block.
\begin{equation*}
\begin{aligned}
    &\forall c\in [c_1, c_2], \forall k \in [0, n_d]: \\
    &\quad \quad \sum_{\mathclap{\substack{i\in(\mathcal{I}\cup \mathcal{B}): c_i=c \\ \Omega(s_i, s_c(i), kt_{\text{day}}, (k+1)t_{\text{day}}) > 0}}} \;
    p_2(\tau_i, \hat{s}_i(kt_{\text{day}}), \hat{e}_i((k+1)t_{\text{day}})) \leq t_{\text{8hr}}
\end{aligned}
\end{equation*}

\noindent \textbf{[POWER-4]:} Imaging within Power Regime 3 is restricted to a maximum of 12 hours per 4-week block.
\begin{equation*}
\begin{aligned}
    &\forall c\in [c_1, c_2], \forall k \in [0, n_d]: \\
    &\quad \quad \sum_{\mathclap{\substack{i\in(\mathcal{I}\cup \mathcal{B}): c_i=c \\ \Omega(s_i, s_c(i), kt_{\text{4wk}}, (k+1)t_{\text{4wk}}) > 0}}} \;
    p_3(\tau_i, \hat{s}_i(kt_{\text{4wk}}), \hat{e}_i((k+1)t_{\text{4wk}})) \leq t_{\text{12hr}}
\end{aligned}
\end{equation*}

\noindent \textbf{[POWER-5]:} A continuous block of at least 8 hours in Power Regime 1 must occur within every 24-hour period to ensure battery recovery.
\begin{multline*}
    \forall c \in [c_1, c_2], \forall t \in [t_0, t_f]: \exists a,b\in[1,|\mathcal{I}|+|\mathcal{B}|]\\ \text{s.t. } \Omega(s_a, s_c(b), t, t+t_{\text{day}}) \geq t_{\text{8hr}}
    \\ \land \forall m \in [a, b]: p_1(\tau_m, s_m, s_c(m)) = s_c(m)-s_m
\end{multline*}

\noindent \textbf{[POWER-6]:} To preserve power for high-draw spacecraft maneuvers, imaging within 8 hours before or after a blackout window is strictly restricted to Power Regime 1.
\begin{equation*}
\begin{aligned}
    &\forall b, c, i: \Omega(s_i, s_c(i), s_b-t_{\text{8hr}}, s_b) > 0 \\
    &\implies p_1(\tau_i, \hat{s}_i(s_b - t_{\text{8hr}}), \hat{e}_i(s_b)) \\
    &\qquad = \Omega(s_i, s_c(i), s_b-t_{\text{8hr}}, s_b)
\end{aligned}
\end{equation*}
\begin{equation*}
\begin{aligned}
    &\forall b, c, i: \Omega(s_i, s_c(i), e_b, e_b+t_{\text{8hr}})>0 \\
    &\implies p_1(\tau_i, \hat{s}_i(e_b), \hat{e}_i(e_b + t_{\text{8hr}})) \\
    &\qquad = \Omega(s_i, s_c(i), e_b, e_b+t_{\text{8hr}})
\end{aligned}
\end{equation*}

\subsection{Scientific and Calibration Constraints}
The full set of science constraints is determined by mission scientists and is extensive. For compactness, we report only representative examples here; these are illustrative of the full requirement set rather than a selective subset. All schedules discussed subsequently in this paper satisfy all science constraints, not only the representative examples shown here.

\noindent \textbf{[SCI-1]:} For every 3 hours of imaging with non-blocked filters, at least one image $\geq 5$ minutes using the blocked filter must be acquired per channel.
\begin{multline*}
    \forall i, j: c_i = c_j \land s_i < e_j \land e_j - s_i \geq t_{\text{3hr}} \\ \implies\exists m \text{ s.t. } \Omega(s_i, e_j, s_m, e_m) > 0 \land f_m = \text{blocked} \\
    \land d_m \geq t_{\text{5min}}
\end{multline*}

\noindent \textbf{[SCI-2]:} The total integration time for each calibration star $k$ on each filter $f$ must meet a required duration $d_{\text{star }k}$ to achieve an SNR $\geq 40$.
\begin{equation*}
    \forall k, f: \sum_{\mathclap{i:\tau_i=\text{star } k \land f_i=f}} d_i \geq d_{\text{star }k}
\end{equation*}

\noindent \textbf{[SCI-3]:} Specific sequences of Earth-pointed images $I_{\text{earth}} = [i_1, \dots, i_m]$ with total duration $d_{\text{earth}}$ must be acquired strictly contiguously at least once per day.
\begin{equation*}
\begin{aligned}
    &\forall k \in [0, n_d]: \exists J_k = [j_1, \dots, j_m] \\
    &\quad \text{ s.t. } \left( \forall x \in [1, m], j_x = i_x \right) \\
    &\quad \land \;\Omega(s_{j_1}, e_{j_m}, kt_{\text{day}}, (k+1)t_{\text{day}}) = d_{\text{earth}} \\
    &\quad \land \left( \forall x \in [1, m-1], s_c(j_x) = s_{j_{x+1}} \right)
\end{aligned}
\end{equation*}

\noindent \textbf{[SCI-4]:} Non-stellar, non-Earth unique calibration sequences indexed by $k$, denoted $I_k=[i_{k_1}, \dots, i_{k_m}]$, must be acquired strictly contiguously exactly once.
\begin{equation*}
\begin{aligned}
    &\forall k: \exists J_k = [j_1, \dots, j_m] \text{ s.t. } \left( \forall x \in [1, m], j_x = i_{k_x} \right) \\
    &\quad\land \;\Omega(s_{j_1}, e_{j_m}, t_0, t_f) = d_k \\
    &\quad\land \left( \forall x \in [1, m-1], s_c(j_x) = s_{j_{x+1}} \right)
\end{aligned}
\end{equation*}

\noindent \textbf{[SCI-5]:} A calibration sequence $I_{\text{cal}} = [i_1, \dots i_m]$ must be scheduled based on cumulative Earth-pointed imaging time. Let $Q = (q_1, \dots, q_m)$ be the ordered list of scheduled $I_{\text{cal}}$ sequences. For convenience, let $q_0 = t_0$ and $q_{m+1} = t_f$. The sequence is required after accumulating 21 hours of Earth imaging, but strictly before exceeding 24 hours, measured from the end of the previous calibration sequence.

\begin{equation*}
    \forall k \in [0, m]: \sum_{i \in E} \Omega(s_i, s_c(i), s_c(q_k), s_{q_{k+1}}) \leq t_{\text{24hr}}
\end{equation*}
\begin{equation*}
    \forall k \in [0, m-1]: \sum_{i \in E} \Omega(s_i, s_c(i), s_c(q_k), s_{q_{k+1}}) \geq t_{\text{21hr}}
\end{equation*}

\subsection{Cost Function}

The cost of a feasible schedule is evaluated by the aggregated responsivity-retrieval error. Let $G$ denote the set of ground-truth responsivities to be validated against, which represent expected on-orbit degradation profiles. For each channel $c$, filter $f$, and ground-truth $g \in G$, we synthetically observe the scheduled stars over $100$ noisy trials (accounting for imperfect stellar spectra knowledge), indexed by $t$, to recover the distribution of responses $\hat{r}_{c,f,g,t}(\lambda)$ \cite{Filippini26}. Let $E_{c,f,g,t}(\lambda) =\|\hat{r}_{c,f,g, t}(\lambda) - r_{c,f,g}(\lambda)\|$ be the absolute retrieval error. We define $\hat{\mu}_{c,f,g}(\lambda)$ and $\hat{\sigma}^2_{c,f,g}(\lambda)$ as the sample mean and sample variance of the relative error, $\frac{E_{c,f,g,t}(\lambda)}{r_{c,f,g}(\lambda)}$, computed across the 100 trials.

The final cost function collates the sample mean error and sample variance across all channels, filters, ground-truths, and the subset of mission-critical wavelengths $\Lambda_f$ given all scheduled images $\mathcal{I}$:
\begin{equation}
\label{eq:cost_function}
    J(\mathcal{I}) = \sum_c \sum_f \sum_{g\in G} \sum_{\lambda \in \Lambda_{f}} \left( \hat{\mu}_{c,f,g}(\lambda) + \frac{1}{4} \hat{\sigma}_{c,f,g}(\lambda) \right)
\end{equation}

\section{Search-Space Analysis}

This section quantifies the combinatorial scale of the scheduling problem defined the previous section and motivates two architectural choices that make our DRL method feasible: (i) activity-block abstraction and (ii) dynamic action masking. For concreteness, we analyze the two-month nominal science commissioning phase (Nov. 10, 2025 to Jan. 10, 2026). A preliminary filter of our calibration star dataset identifies $321$ stars visible during this period, which is more than can be feasibly observed within the available time.

\subsection{Atomic Scheduling vs. Activity Blocks}

In order to apply DRL, we formulate an atomic Markov Decision Process (MDP). Each exposure is an independent decision. Therefore, the instantaneous action space consists of targets (321 stellar targets + 18 calibration targets + Earth + “no target”), filters (one per channel), and integration times, yielding $a = 341 \times 36 \times 60 \approx 7.37 \times 10^5$ actions. Over a 60-day horizon requiring $\sim 4,300$ decisions (see Table \ref{tab:science_quality_two} for the number of images in example schedules and note that this is conservative), the search tree spans roughly $10^{25,000}$ permutations. This scale, heavily burdened by long-range temporal dependencies and a sparse feasibility manifold, is intractable for classical heuristic search, constraint programming, and vanilla model-free RL.

\begin{figure*}[t]
\centering
\includegraphics[width=0.7\textwidth]{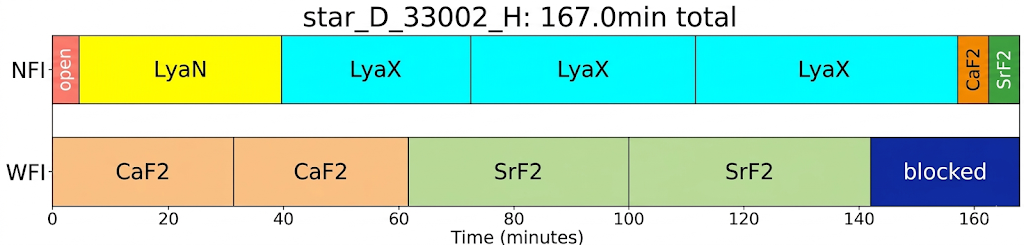}
\caption{Example activity block with a single stellar target. Each box represents an image; rows correspond to channels and labels to filters.}
\label{fig:activity_block_example}
\end{figure*}

To make learning tractable, we lift the action space using \emph{activity blocks}, defined as macro-actions that aggregate atomic exposures into mission-validated observation sequences. Using the notation from Section \ref{sec:sched_prob}, activity blocks are defined as $\mathcal{A}=(\mathcal{I},c_a, s_a)$, where $\mathcal{I}$ is a sequence of images such that $\forall i \in \mathcal{I}:c_i=c_a$ and each $s_i$ is the start time of image $i$ given that the activity block starts at $s_a$.

Activity blocks can be designed to meet constraints INST-1, INST-2, INST-3, INST-5, INST-6, INST-7, and SCI-1 individually, so that a schedule consisting of non-overlapping activity blocks on each channel can also satisfy these constraint globally. Some block designs are strictly dictated by science constraints (such as the image sequences in SCI-3, SCI-4, and SCI-5) and are thus statically defined, while other blocks are deterministically generated. For example, the pipeline to generate stellar activity blocks calculates the required integration times for a Signal-to-Noise Ratio (SNR) of $40$, splits exposures to respect detector limits, and optimizes filter ordering for every candidate star independently. Figure \ref{fig:activity_block_example} shows an example of a stellar activity block. Activity blocks are implemented using a Python class to ensure that all blocks, whether defined manually by scientists or generated via custom code, adhere to a unified interface for constraint checking.

Operating at the block level shrinks the horizon to roughly $900$ decisions (see Table \ref{tab:science_quality_two}), which reduces the search space to $342^{900} \approx 10^{2,280}$. Beyond algorithmic tractability, this abstraction enhances operational trust by factorizing constraint checking into two independent checks: one for the library of activity blocks, and one for the final schedule. The abstraction also improves agility, as activity blocks are straightforward to adjust if science priorities change. While the specific activity blocks are unique to this mission, the methodology of encapsulating risky atomic constraints into safe macro-actions is a generalizable and necessary precursor for applying RL to real-world operations.

\subsection{Dynamic Action-Masking}

Despite the macro-action abstraction, the vast number of activity block sequences violate global operational constraints (e.g., daily Earth-pointing, solar-angle limits, and power-recharge windows). Empirically, we observe that $\sim 90\%$ of nominal actions are dynamically infeasible at any given timestep.

We mitigate this issue by using dynamic action-masking. Invalid blocks are assigned large negative logits, reducing their selection probability to zero. This constraint-enforcement layer shrinks the effective search space to roughly $(0.1 \times 342)^{900} \approx 10^{1,380}$, as all schedules now automatically satisfy INST-8 and all POWER constraints. Action masking transforms the learning problem from feasibility-seeking to science-optimization within a valid manifold.

\section{DRL-Based Scheduling Architecture}
\label{sec:rl_sched_sol}

\begin{figure*}
    \centering
    \includegraphics[width=1\linewidth]{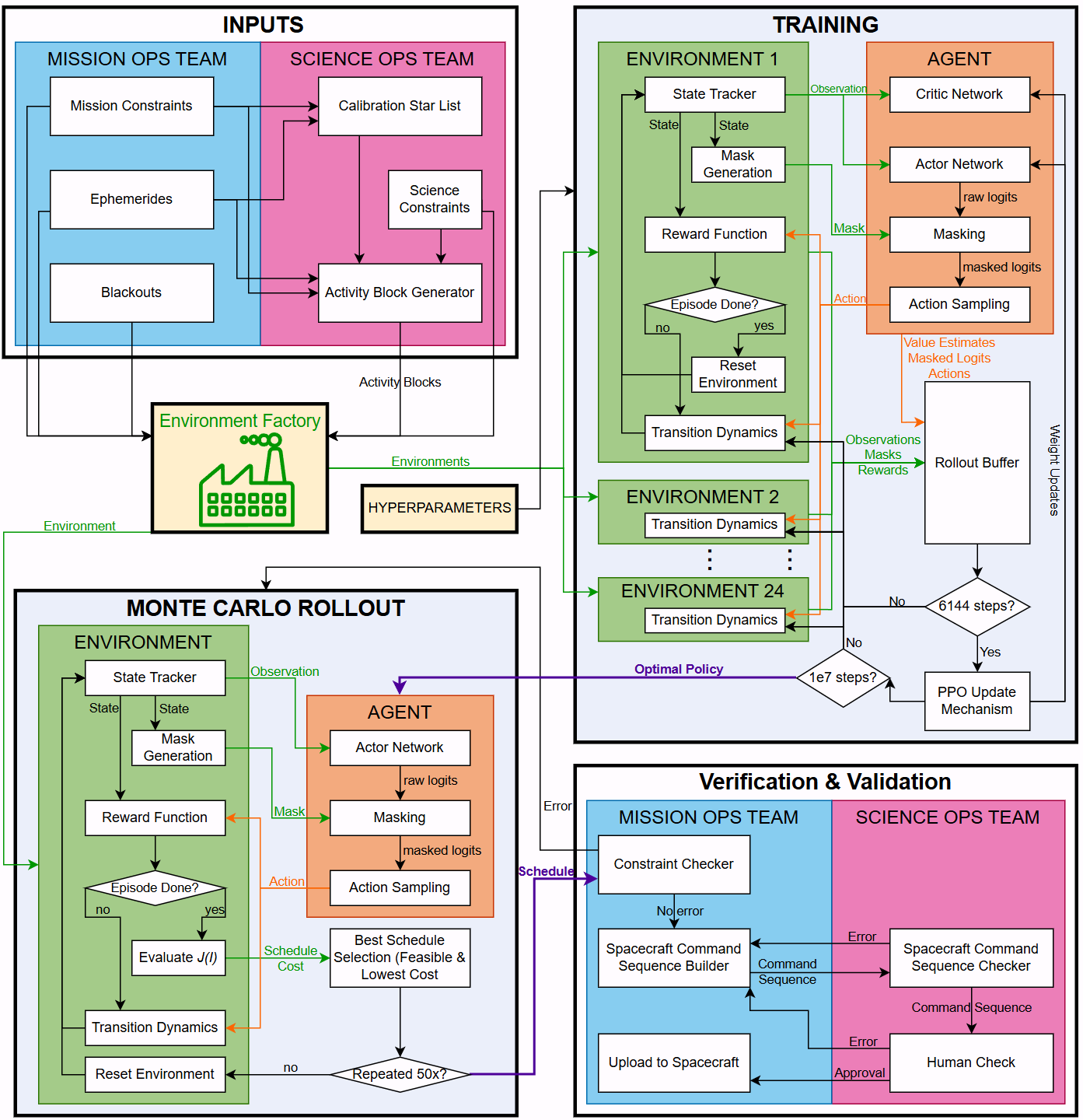}
    \caption{Deep Reinforcement Scheduler Architecture. The mission operations team and the science operations team's respective inputs to the scheduler are shown in the top-left box. The environment factory uses these inputs to create environments. The DRL training loop is shown in the top-right box, with details found in Section \ref{sec:rl_sched_sol}. After training, the final schedule is generated by the optimal policy using Monte Carlo rollout (bottom-left) to obtain a feasible schedule with overwhelming probability. The schedule is then validated by both the mission operations team and the science operations team independently to ensure all constraints are met before upload to the spacecraft.}
    \label{fig:architecture}
\end{figure*}

In this section, we summarize the operational RL formulation and implementation decisions used to solve the scheduling problem faced by the Carruthers mission. A diagram of the architecture is shown in Figure \ref{fig:architecture}. We implemented a custom environment for this scheduling problem using the OpenAI Gymnasium interface \cite{towers2024gymnasiumapi}. We discretize time into $30$-second steps, since all activity blocks and blackout zones are guaranteed to be multiples of $30$ seconds due to instrument characteristics. 

The state space is defined by a flat vector of $359$ variables, normalized to $[0, 1]$ to stabilize training. A single observation consists of the simulation clock (in $30$-second intervals from $t_0$), remaining time in Power Regimes 2 and 3, recovery metrics (time spent in Regime 1 over the last $24$ hours and time since the last Regime 1 image), stellar observation counts per filter, cumulative Earth-imaging time (critical for evaluating the SCI-5 constraint), and the execution counts of each activity block.

The dynamic masking function is evaluated at each decision step, where activity blocks are masked if their execution would violate the single-execution constraint of SCI-4, any of the thermal and power limits (POWER-1 through POWER-6), or if the proposed activity block is occluded by blackout regions (INST-8). We include a do-nothing action, which simply steps the simulation clock forward by $30$-seconds, to ensure that there is always an action available.

We utilize reward shaping to guide the policy toward feasible, high-quality schedules while mitigating sparse-reward challenges. We apply positive rewards for successfully scheduling mandatory sequences (SCI-3 and SCI-4), coupled with heavy penalties if the horizon ends without their completion or if Earth-imaging constraints (SCI-5) are violated.

The reward structure for science quality is considerably more complex. Evaluating the schedule's cost function $J(\mathcal{I})$ (see Equation \ref{eq:cost_function}) once at the end of an episode was found to be too sparse. However, evaluating $J(\mathcal{I})$ after every decision step where a star was scheduled is prohibitively expensive due to the number of ground-truths and trials involved. Consequently, we execute a partial validation every two simulated days. A channel/filter combination is only evaluated once the policy has scheduled at least one star for that combination, and only a single ground-truth $g$ (randomly selected from $G$) is tested in each partial validation. On the first such evaluation, the reward is a penalty similar to $J(\mathcal{I})$:
$$J'(\mathcal{I})=\sum_c \sum_f \sum_{\lambda \in \Lambda_f}\left(\hat{\mu}_{c,f,g}(\lambda) + \frac{1}{4} \hat{\sigma}_{c,f,g}(\lambda)\right)$$

On subsequent evaluations, the reward is the improvement (or deterioration) on $J'(\mathcal{I})$, scaled by a reward factor $\alpha = 29(1 - \cos(\pi \theta^2))/2 + 1$, which increases from $1$ to $30$ following a squared-cosine schedule. Here, $\theta$ is the fraction of time that has already passed in the full scheduling horizon. The scaling is designed such that late-episode improvements, which are typically smaller because many stars have already been observed, are valued similarly to the larger early-episode gains when only a few stars have been observed. Furthermore, we found that rewarding relative improvement, rather than strictly penalizing absolute error, is critical for sustained exploration. Under a purely penalty-based regime, the policy quickly collapses into a stalling behavior: because activity blocks (actions) advance the simulation clock by varying durations, the agent learns to repeatedly select the shortest available action (the 30-second do-nothing action) simply to defer impending negative rewards as long as possible. To directly combat this exploit and promote schedule density, we additionally impose a small baseline penalty whenever the do-nothing action is selected.
 
We train our policy using the Stable Baselines3 \cite{stablebaselines3} implementation of Maskable Proximal Policy Optimization (MPPO) \cite{mppo2020}, which extends vanilla PPO \cite{ppo2017} to support dynamic action masking (see the Training box in Figure \ref{fig:architecture}). To accelerate data collection, we vectorize $24$ independent environment instances, aggregating 6,144 decision steps per model update. The policy and value functions are parameterized by standard multi-layer perceptrons (a two-hidden-layer $128$-$128$ architecture for the policy network and a three-hidden-layer $256$-$256$-$128$ architecture for the value network). Training proceeds for $10^7$ total timesteps, which consistently yields a robust stochastic policy. We found that the average episode length decreases from roughly 2,000 to 1,700 steps as the policy learns to efficiently navigate blackout regions and avoid the do-nothing action.

After training completes, we obtain the final schedule by completing $50$ full episodes using the trained stochastic policy and evaluate $J(\mathcal{I})$ for each feasible schedule (see the Monte Carlo Rollout box in Figure \ref{fig:activity_block_example}). Empirically, it is exceedingly rare for any of these $50$ rollouts to produce an infeasible schedule. The final schedule is selected as the feasible schedule that minimizes $J(\mathcal{I})$.

\section{Science-Quality Performance}

To assess the science quality of the schedules produced by the learned policy, we compare them against schedules generated by three lightweight heuristics. All heuristics employ activity blocks for easy comparison. Each heuristic constructs a full schedule by first placing all required observations as soon as they become available: the daily Earth-pointed imaging (SCI-3), the fixed non-stellar observations (SCI-4), and the special calibration images (SCI-5). However, this initial step was not able to schedule all required activities, so they were manually adjusted to ensure full feasibility. Multiple hand-crafted variants were then produced by perturbing the placement of required observations. Any remaining gaps were filled with stellar observations selected according to the heuristic being evaluated. For each gap of time, two candidate star sets were defined: a do-no-harm set consisting only of stars in Power Regime 1 and a feasible set containing all stars that satisfy all pointing and power constraint. Before the final required observation, only do-no-harm stars were allowed to be scheduled; afterward, all feasible stars became eligible. The three heuristics described below determine how to select from these candidate sets.

The baseline heuristic selects a star uniformly at random from the currently available candidate set to schedule. This gives a simple, feasibility-preserving reference schedule without any reasoning about stellar targets.

To construct stronger heuristics, we incorporate optimal experimental design (OED) principles \cite{oed_formulation}. OED treats data collection as the problem of choosing the next experiment to maximize the expected reduction in uncertainty of some underlying estimation problem. In our case, the estimation problem is recovering the instrument’s responsivity curve from stellar photometric measurements for each channel/filter combination. We wish to minimize the uncertainty of the instrument's responsivity, which also minimizes $J(\mathcal{I})$ (see Equation \ref{eq:cost_function}). For a candidate star $i$, we compute the posterior covariance matrix $\Sigma_{\text{post}}$ by updating the prior covariance with the star's Fisher information matrix, which is derived from its known photon flux and noise level. Our two OED heuristics rank available stars by scoring this updated covariance: the A-optimality heuristic minimizes the sum of posterior variances ($\text{tr}(\Sigma_{\text{post}})$), favoring stars that reduce average uncertainty across all $\lambda \in \Lambda_f$, while the D-optimality heuristic minimizes the posterior uncertainty volume ($\text{det}(\Sigma_{\text{post}})$) (i.e., maximizes information gain).

Tables \ref{tab:science_quality_one} and \ref{tab:science_quality_two} report the highest-quality instance obtained for each heuristic and the schedule obtained by the trained DRL policy over one- and two-month horizons. Across both horizons, the DRL policy schedules a comparable number of stars while achieving the lowest overall cost, outperforming all heuristics. Notably, the DRL approach requires less total integration time on calibration stars than the baseline heuristics, which directly translates to increased temporal allocation for primary mission science.

\begin{table*}[t]
\centering
\begin{tabular}{|l|c|c|c|c|c|}
    \hline
    \multicolumn{6}{|c|}{\textbf{One-month horizon}} \\
    \hline
    Method & Cost & \# Stars & \% Time on Stars & \# Activity Blocks & \# Images \\
    \hline
    Baseline heuristic & 289.56 & 157 & 52.0 & 335 & 2145 \\
    A-optimality heuristic & 268.03 & 157 & 51.1 & 346 & 2170 \\
    D-optimality heuristic & 283.31 & 145 & 52.9 & 320 & 2078 \\
    \textbf{DRL} & \textbf{256.98} & 149 & \textbf{43.8} & 377 & 2204 \\
    \hline
\end{tabular}
\caption{Science-quality performance across scheduling methods for a one-month horizon (Nov. 10th 2025 to Dec. 10th 2025). The cost reported in column 2 is $J(\mathcal{I})$ (see Equation \ref{eq:cost_function}). Column 3 reports the number of stars observed, while column 4 reports the percentage of the full horizon spent observing stars. Since the Carruthers mission is an Earth-observing mission, less time spent on calibration stars directly implies more time spent on primary science and is thus desired. The last columns indicate the number of activity blocks and images in the best schedule produced by each method.}
\label{tab:science_quality_one}
\end{table*}

\begin{table*}[t]
\centering
\begin{tabular}{|l|c|c|c|c|c|}
    \hline
    \multicolumn{6}{|c|}{\textbf{Two-month horizon}} \\
    \hline
    Method & Cost & \# Stars & \% Time on Stars & \# Activity Blocks & \# Images \\
    \hline
    Baseline heuristic & 270.48 & 223 & 36.9 & 819 & 4334 \\
    A-optimality heuristic & 260.41 & 219 & 35.7 & 847 & 4346 \\
    D-optimality heuristic & 252.64 & 225 & 37.2 & 825 & 4343 \\
    \textbf{DRL} & \textbf{189.94} & 202 & \textbf{30.7} & 720 & 4349 \\
    \hline
\end{tabular}
\caption{Science-quality performance across scheduling methods for a two-month horizon (Nov. 10th 2025 to Jan. 10th 2026). See caption of Table \ref{tab:science_quality_one} for column descriptions.}
\label{tab:science_quality_two}
\end{table*}

\section{Operational Performance}

We now evaluate the DRL framework's operational performance by systematically revisiting the three primary impediments to DRL-based scheduling identified in the introduction: operational trust, agility, and strict feasibility. This section discusses how our approach resolves each in practice.

\subsection{Operational Trust}

Verifiable safety was central to the system's adoption by both mission and science operations teams. By abstracting atomic images into activity blocks, we effectively decoupled the constraint-verification process. Validating a monolithic, image-level schedule requires evaluating all micro- and macro-constraints simultaneously. In contrast, our framework allows for a two-tier verification: operators first independently certify that the static activity blocks satisfy all local instrument constraints, and then independently verify that the global sequence of activity blocks respects spacecraft-level power and thermal limits. This factorization drastically reduces the complexity of day-to-day schedule validation.

Plan comprehensibility was equally critical for the science operations team. Representing actions as activity blocks produced schedules whose structure aligned with established operational concepts. During internal mission reviews, mission scientists consistently noted that this macro-level representation was far easier to parse than raw image-level sequences, significantly reducing the cognitive burden of day-to-day science constraint validation.

\subsection{Agility}

On a workstation with an AMD Threadripper 5995WX and an NVIDIA RTX 4070, end-to-end training requires approximately five hours, while the Monte Carlo rollout requires approximately thirty minutes. These short execution times provide same-day scheduling capabilities suitable for both routine operations and contingency response. Crucially, this computational efficiency circumvents the traditional challenge of policy robustness to shifting constraints or science priorities; rather than requiring a hyper-generalized policy to accommodate unknown potential changes, our framework permits a full, on-demand retrain.

The system has been demonstrated to support time-critical anomaly responses. During a pre-launch operational readiness test, a large blackout window associated with a spacecraft maneuver was shifted by three days, invalidating the existing multi-week plan. The science team had twelve hours to produce a revised schedule that met all imaging-based science constraints and respected the new blackout interval while still optimizing for science quality. After adjusting only the blackout-time mask, the RL framework generated a new conflict-free schedule in roughly six hours, with no manual adjustments required.

This computational efficiency has enabled rapid re-planning during periods of constraint volatility. As is typical for all missions approaching launch, operational constraints evolved repeatedly during integration and testing: modifications to power budgets, permissible exposure durations, and time-dependent pointing windows all occurred within weeks of delivery as both spacecraft and science requirements solidified. Traditional heuristics, like the ones presented in this paper, would require re-authoring scheduling logic and manual adjustments whenever such constraints shifted. In contrast, the RL-based framework required only updates to specific activity blocks or the action-masking specification, with retraining automatically adapting to any changes without requiring any hyperparameter retuning. This agility was operationally valuable: constraint updates that previously required days of human effort and were prone to human error could be reliably absorbed into the scheduling pipeline within hours.

\subsection{Strict Feasibility}

Across all experiments conducted after hyperparameter tuning, 99\% of generated schedules were fully feasible without requiring any manual intervention or post-processing. The small fraction of infeasible cases were attributable entirely to missing some of the required imaging mandated by the SCI-3, SCI-4, or SCI-5 constraints. Notably, because dynamic action-masking mathematically precludes any violation of operational limits, these edge cases only compromise scientific objectives while maintaining absolute guarantees on spacecraft health. In practice, the high feasibility rate implies that the Monte Carlo rollout yields a globally feasible schedule with overwhelming probability, entirely eliminating the need for manual constraint repair.

Ablation studies confirm that vanilla PPO consistently fails to find feasible schedules within the same training budget. Activity-block abstraction and dynamic masking are therefore not mere performance optimizations; they are mandatory architectural choices for solving this constrained scheduling problem with DRL.

\subsection{On-orbit Operations}

The DRL framework has transitioned from pre-launch testing into active flight operations after approval by NASA and has served as the default operational scheduler for the Carruthers mission since launch. The combination of operational trust, agility, and strict feasibility has enabled the framework to serve not merely as a research prototype but as a deployed operational asset for NASA science operations. To our knowledge, this represents the first application of DRL as the default operations scheduler for a large-scale NASA mission. 

\section{Conclusion}

In this work, we presented a scalable deep-reinforcement-learning framework for long-horizon spacecraft operations scheduling and demonstrated its successful application to the Carruthers Geocorona Observatory mission. By combining two key architectural choices, namely activity-block macro-actions and dynamic action-masking, we reduced the effective search space by several orders of magnitude and enabled stable policy training for a problem far beyond the reach of classical methods. This combination embeds operational knowledge directly into the decision structure, enforces feasibility throughout training, and yields schedules with enhanced science quality.

The DRL framework described in this paper deals with the three major issues that most DRL scheduling solutions struggle with. Activity blocks allow constraint checks to be factorized and improve comprehensibility of plans, thereby gaining operational trust. Only six hours of computation are required for training and Monte Carlo rollout, allowing high agility in response to constraint and/or science priority changes. Finally, action-masking and activity block design allow the learned policy to consistently produces globally feasible schedules that outperform OED-based heuristic baselines.

The scheduling framework was adopted as the default operational scheduler from the start of the Carruthers mission and has been successfully driving flight operations for the six continuous months since mission launch at the time of writing. This is a significant milestone, demonstrating that DRL methods can be trusted for critical, real-world mission operations when paired with appropriate semantic abstractions and rigorous constraint-handling mechanisms. While our study focuses on a single mission, the underlying methodology of lifting the action space through validated macro-actions and enforcing feasibility via dynamic masking offers a practical, reusable template for other spacecraft scheduling scenarios. With the right architectural alignment to human operational workflows, DRL can be made highly reliable, agile, and operationally vital, even in domains characterized by stringent constraints and long planning horizons that require high agility.

\appendix

\section{Code Disclaimer}

Due to mission restrictions, the code is currently proprietary. We expect portions to be released at a later date; meanwhile, contact the first author with inquiries.

\bigskip
\bibliography{aaai2026}


\end{document}